\documentclass[fleqn,usenatbib,referee]{mnras}

\usepackage{newtxtext,newtxmath}

\usepackage[T1]{fontenc}
\DeclareRobustCommand{\VAN}[3]{#2}
\let\VANthebibliography\thebibliography
\def\thebibliography{\DeclareRobustCommand{\VAN}[3]{##3}\VANthebibliography}

\usepackage{graphicx}	
\usepackage{amsmath}	
\usepackage{amssymb}	

\title[A dynamo model for $\varepsilon$ Eridani]{An application of a solar-type dynamo model for $\varepsilon$ Eridani}

\author[A. P. Buccino et al.]{
A. P. Buccino,$^{1,2}$\thanks{E-mail: abuccino@iafe.uba.ar}
L. Sraibman,$^{3}$
P. M. Olivar$^{1,3}$
and F. O. Minotti$^{1,3}$
\\
$^{1}$ Departamento de F\'\i sica, FCEyN Universidad de Buenos Aires, Buenos Aires, Argentina. \\
$^{2}$ Instituto de Astronom\'\i a y F\'\i sica del Espacio (UBA-CONICET), Buenos Aires, Argentina.\\
$^{3}$ Instituto de F\'\i sica del Plasma (UBA-CONICET), Buenos Aires, Argentina.
}

\date{Accepted XXX. Received YYY; in original form ZZZ}

\pubyear{2020}

\begin{document}
\maketitle

\begin{abstract}
During the last decade, the relation between activity cycle periods with stellar parameters has received special attention. The construction of reliable registries of activity reveals that solar type stars exhibit activity cycles with periods from  few years to decades and, in same cases, long and short activity cycles coexist suggesting that two dynamos could operate in these stars.
In particular, $\varepsilon$ Eridani is an active young K2V star (0.8 Gyr), which exhibits a short and long-term chromospheric
cycles of $\sim$3 and $\sim$13-yr periods. Additionally, between 1985 and 1992, the star went through a broad activity minimum, similar
to the solar Maunder Minimum-state.

Motivated by these results, we found in $\varepsilon$ Eridani a great opportunity to test the  dynamo theory. Based on the model developed in \cite{Sraibman19}, in this work we built a non linear axisymmetric  dynamo for $\varepsilon$ Eridani.  The time series of the  simulated  magnetic field components near the surface   integrated in all the stellar disc exhibits both the long and short-activity cycles with periods similar to the ones detected from observations and also time intervals of low activity which could be associated to the broad Minimun. The short activity cycle associated to the magnetic reversal could be explained by the differential rotation, while the long cycle is associated to the meridional mass flows induced by the Lorentz force. In this way, we show that a single non-linear dynamo model derived from first principles with accurate stellar parameters could reproduce coexisting activity cycles. 
\end{abstract}




\section{Introduction}\label{sec.intro}

Stellar activity cycles, similar to the solar one, have been observed in  several solar-type stars \cite{Baliunas95,Messina02,GomesdaSilva14} and even in few dM stars \cite{Diaz07,Buccino11,Buccino14,Ibanez19}.

The $\alpha\Omega$ dynamo is the most accepted theory to model the mean magnetic field in cyclic late F to early M stars (e.g. \cite{Durney82,Lorente05}). Driven by differential
rotation ($\Omega$-effect) and turbulent convection ($\alpha$-effect) in the stellar interior, this model predicts a strong correlation between activity and rotation. The  $\alpha\Omega$ dynamo was first invoked to model solar magnetic activity (e.g. \cite{Parker55,Parker63, Steenbeck69}). However, the physical origin of solar activity is more complex than the one suggested by these earlier models (e.g. \cite{Choudhuri95,Dikipati99,Jouve07,Rempel06,Passos2014,Passos2015,Miesch2016, Hazra2017,Lemerle2017}). 



The Mount Wilson survey has provided relevant works which document the relation between  stellar activity: level of activity, cycle period, etc. and fundamental stellar parameters: mass, rotation period, luminosity, age, convection, etc. (e.g. \cite{Noyes84,Soderblom91,Soon93,Donahue96,Brandenburg98,Saar99}). In particular, \cite{Brandenburg98} selected a set of cyclic stars  with well-determined rotation and activity periods. They explored the non-dimensional relationships between the activity cycle period $P_{cyc}$, the rotational period $P_{rot}$ with different parameters. These relations reveal two different sequences, called the \textit{A-branch}, as primarily \textit{active} stars belong to this branch, and the \textit{I-branch} for the \textit{inactive} stars which crowd this sequence. Both parallel branches with positive slopes are separated by a factor of $\sim$6 in $P_{cyc}/P_{rot}$ diagrams. 
Furthermore, \cite{Saar99} expanded this previous work for further active and older stars. They confirmed the bimodal distribution observed in \cite{Brandenburg98} for stars older than 0.1 Gyr and detected that the most active stars  populate a third branch with negative slope, called \textit{superactive}. In particular, they detected that a set of young stars older than 1 Gyr of the active and inactive  group present activity cycles in both sequences. The long cycle falls on the A-branch, while the short cycle, considered as a secondary cycle, on the inactive (I) branch. 
A possible interpretation to this bimodal distribution is that the $\alpha$-effect is an increasing function of magnetic
field strength on the active and inactive branches (known as \textit{anti-quenching} $\alpha$-effect), while for larger magnetic fields the quenching effect is present as observed in the superactive branch \cite{Saar99}. 

\cite{BohmVitense07} explored the relation between $P_{cyc}$ and $P_{rot}$ for a set of dF and dG of the sample analyzed in \cite{Saar99} and \cite{Lorente05}. The advantage of the $P_{cyc}$-$P_{rot}$ diagram is that it only depends on observational measurements, independent of stellar models.  Two sequences (\textit{active} and \textit{inactive}) as first suggested in \cite{Brandenburg98} are evident in this diagram. A tentative interpretation  provided by \cite{BohmVitense07} is that a dynamo operating near the surface
could be generating the longer cycle, while a  a second
  dynamo operating in the deep convection zone could be responsible
  for the shorter one. She also suggests that both dynamos could work simultaneously, responsible for the  co-existing long and short activity cycles detected in several stars \cite{Saar99,Metcalfe13,Flores17,Egeland15}. \cite{Strugarek18} addresses the two-branches subject with stellar dynamo models. They  performed 3D MHD simulations  varying the rotation rate and luminosity of a set of modeled solar-like convective envelopes. They detected both primary and secondary cycles on experiments, when the local Rossby number ($R_o$) is lower than 1. However, recently \cite{Boro18} expanded the sample analyzed by \cite{BohmVitense07}, including stars with less precisely determined activity cycles. They concluded that stars can lie in the region intermediate between the active and inactive branch, putting in doubt a bimodal distribution.

  Furthermore,  \cite{BohmVitense07}'s work brings up the controversial fact that the Sun lies between both branches in $P_{cyc}-P_{rot}$ diagram.  Recently, \cite{Brandenburg17} compiled all new and previous observations of stellar activity  and  confirmed that long and short
stellar cycles tend to fall on one of two universal branches and concluded that \emph{all} stars younger than 2.3 Gyr are capable of exhibiting simultaneously longer and shorter cycle periods. Furthermore, the Sun does not lie between this new
two branches.

To interpret the solar activity in a stellar context, \cite{vanSaders16} analyzed \textit{Kepler} photometric data which allow them to build a reliable  rotation-age relation for stars older than the Sun, where they found that beyond middle-age the efficiency of magnetic braking is dramatically reduced. Combining these data with chromospheric activity indicators, \cite{Metcalfe16} show that the activity level continues to decrease with a smooth evolution of the rotation rate. Therefore, they conclude that the Sun
could be in a transitional evolutionary phase which place it between both active and inactive branches. They suggest that the Sun might represent a special case of stellar dynamo theory. However, \cite{Strugarek17} normalized \cite{BohmVitense07}'s diagram with the stellar luminosity and observed that the Sun fit the same trend that solar-type stars. Therefore, these observations reinstate the Sun to the status of an ordinary solar-type star. The discussion is still open, without 
certainties  on whether the Sun is in a transitional phase or not.


In particular, $\varepsilon$ Eridani (HD 22049) is an active young K2V star (0.8 Gyr), which has been characterized as an exoplanet host star (e.g. \cite{AngladaEscude12}).  $\varepsilon$ Eridani is one of the nearest and brightest stars in the sky, extensively observed, thus it provides an opportunity to constrain stellar dynamo theory. In this sense,  \cite{Metcalfe13} built a  registry of activity of 45-year span for this star by combining the Mount Wilson indexes obtained from  different observatories.  This time-series is a unique registry of stellar activity, beyond the Sun. It shows two coexisting 3-year and 13-year activity cycles and a  broad activity minimum between 1985 and 1992 that resembles a Maunder minimum-like state. Both activity cycle periods detected belong to the \textit{active} and \textit{inactive} branches of Bohm-Vitense's diagram. All these facts make $\varepsilon$ Eridani an ideal target to test the dynamo theory.

To take a different approach to this discussion, we apply the dynamo model developed in \cite{Sraibman19} to the active cool star $\varepsilon$ Eridani and we analyzed if a solar-type dynamo model could reproduce both activity cycles reported by \cite{Metcalfe13}. 

In section \S\ref{sec.model}, we review the main characteristic of the dynamo model employed in this work. In section \S\ref{sec.results}, we show the magnetic field maps derived from this model and, also, the evolution of the $\alpha$ coefficient and the induced meridional velocity field. In section \S\ref{sec.discuss} we analyze the long-term magnetic activity of this star through a magnetic activity proxy  defined in this work and summarize our main conclusions.

\section{A non-linear dynamo model}\label{sec.model}


Here we consider a 2-D axisymmetric dynamo model coupled with dynamic equations describing the back reaction of the magnetic field on the mass flow. The model, developed, and tested for the case of the Sun, in \cite{Sraibman16,Sraibman19}, determines the $\alpha$ and diffusivity tensors in terms of the large-scale fields, using a novel technique for deriving large-scale equations directly from the original equations, employing only first principles (\cite{Minotti2000}). This technique leads to the separation of the problem into three scales: large scales resolved by the numerical procedure, intermediate scales whose dynamic effect on the resolved scales is modeled by the mentioned technique, and microscales that are described separately by simpler models.

The stellar parameters of $ \varepsilon $ Eridani were obtained from \cite{Gai08}. These data were employed in the stellar model developed by \cite{Eggleton71} to compute the density in the stellar interior, the tachocline depth and width. To do so, we employed the public code \textsf{EZ Web}\footnote{http://www.astro.wisc.edu/~townsend/}.

One of the main ingredients of our model is the parametrization of the differential rotation. In particular, the surface differential rotation period of  $\varepsilon$ Eridani $P_{\theta_1}$ has already been determined from \textit{MOST}\footnote{Microvariability and Oscillations
of STars microsatellite} photometric data  in \cite{Cr6}:

\begin{equation}
P_{\theta_1}=\frac{P_{EQ}}{1-k.sin^{2}\theta_1 }\label{eq.ptheta}
\end{equation}\\
where $P_{EQ}$= 11.20 days, $k $= $0.11$ and $\theta_1$ is the latitude.

\begin{figure}
  \centering
    \includegraphics[width=0.6\textwidth]{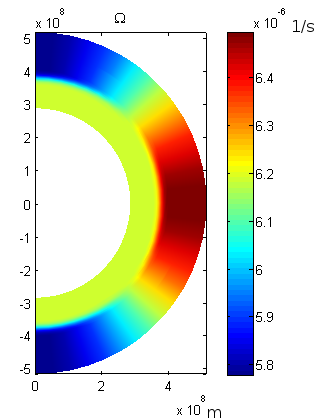}
  \caption{$\Omega$-profile in the convective layer given by Eq. \ref{eq.omega}. }
  \label{fig.omega}
\end{figure}

 We assume a solar-type differential rotation in the convective layer given by: 
\begin{equation}
\Omega(r,\theta) =\Omega _{RZ} + 0.5\left(1+erf\left(\frac{2(r-rt)}{dt}\right)\right)\ast   \left( \frac{2\pi }{P_{\theta_1}}-\Omega_{RZ}\right) \end{equation}
\label{eq.omega}\\
where $\Omega_{RZ}$ is the rotation of the core ($\Omega_{RZ}=2\pi 9.84\times 10^{-7} \mathrm{s}^{-1}$), $dt$ is the tachocline width($dt$=0.05$rt$),
$rt=0.7276R_{\star}$ is the tachocline radius in $\varepsilon$ Eridani (see Fig. \ref{fig.omega}) and $erf(x)$ is the error function.

The tachocline radius $rt$  and its width $dt$ were derived with the \cite{Eggleton71}'s model, these  values are closed to ones derived by \cite{Gai08} from asteroseismic observations. The rotation of the core $\Omega_{RZ}$ is proportional to the solar one $\Omega_{RZ}^\odot$ considering the relation between the solar and stellar rotation periods and radii. 

 In \cite{Sraibman17}, we first built a kinematic dynamo model with the parameters above for $\varepsilon$ Eridani based on \cite{Sraibman16}, but we could only reproduce the short cycle. In the present work we adopt the model developed in \cite{Sraibman19}, where magnetic feedback on the mass flow  was included.

In \cite{Sraibman17} the $\alpha$ coefficient was derived in terms of the radial cylindrical component of the mean vorticity $\omega_s$: 
 \begin{equation}
     \alpha=\varkappa \frac{\lambda ^{2}}{24\,s}\omega_s 
 \end{equation}
where $s=r sin\theta$, $\theta$ is spherical polar angle, $\lambda$ is the large-scale length  ($\lambda=0.05R_*$ for $\varepsilon$ Eridani) and $\varkappa$ is an adjustable constant between 0 and 1. 
 
 In  \cite{Sraibman19}  the $\alpha$ coefficient is composed of two main additive contributions $\alpha^{(0)}$ and $\alpha^{(1)}$:
\begin{equation}
     \alpha^{(0)}=\varkappa \frac{\lambda ^{2}}{24\,s}\omega_s^{(0)} 
     \label{eq.alf0}
 \end{equation}
 where the base contribution to $\omega_s$ comes from the differential rotation, for the axisymmetric case considered:
 
 \begin{equation}
     \omega^{(0)}=sin\theta[sin\theta\frac{\partial \Omega}{\partial\theta}-rcos\theta\frac{\partial \Omega}{\partial r}]
     \label{eq.omega0}
     \end{equation} 
 with $\Omega(r,\theta)$ given by Eq. \ref{eq.omega}.
 
 The $\alpha^{(1)}$ term is related to the vorticity component $\omega_s^{(1)}$ of the flow generated by the feedback effect:
 \begin{equation}
     \alpha^{(1)}=\varkappa \frac{\lambda ^{2}}{24\,s}\omega_s^{(1)},
     \label{eq.alf1}
     \end{equation}
 with $\omega_s^{(1)}$  given by
  \begin{eqnarray*}
    \frac{\partial\omega_s^{(1)}}{\partial t}+U_s^{(0)}\Bigg (\frac{\partial\omega_s^{(1)}}{\partial s} + \frac{\omega_s^{(1)}}{s}\Bigg)+U_z^{(0)}\frac{\partial\omega_s^{(1)}}{\partial z} +\\
     \omega_s^{(1)} \frac{\partial U_z^{(0)}}{\partial z}=2\Omega_0\frac{\partial U_s^{(1)}}{\partial z}-\omega_s^{(0)}\frac{\partial U_z^{(1)}}{\partial z}-\\
U_s^{(1)}\Bigg (\frac{\partial\omega_s^{(0)}}{\partial s} + \frac{\omega_s^{(+)}}{s}\Bigg)-U_z^{(1)}\frac{\partial\omega_s^{(0)}}{\partial z}
     \end{eqnarray*}

where the $z$-axis coincides with the direction of the stellar rotational axis,  $U_{s}$ and $U_z$ are the average velocity field components in the radial cylindrical direction  and the z-direction respectively. 
\section{Results}\label{sec.results}

We perform a simulation of more than 1500 yrs starting with zero meridional velocity and a small dipolar magnetic field (0.01 nT). We discard the  first 200 yrs of simulation during which the magnetic field reaches the permanent regime. 

\begin{figure}
\centering
    \includegraphics[width=\textwidth]{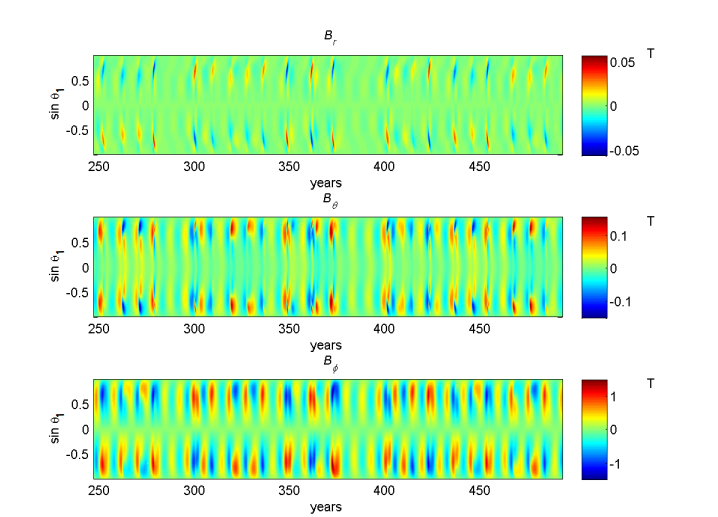}
  \caption{Magnetic field components of $\varepsilon$ Eridani as a function of time and latitude just below the surface, at r=0.98$R_*$. The
top and middle panels show the poloidal components  and  the bottom panel presents the toroidal component.}
  \label{fig:btime}
\end{figure}

\begin{figure}
     \includegraphics[width=\textwidth]{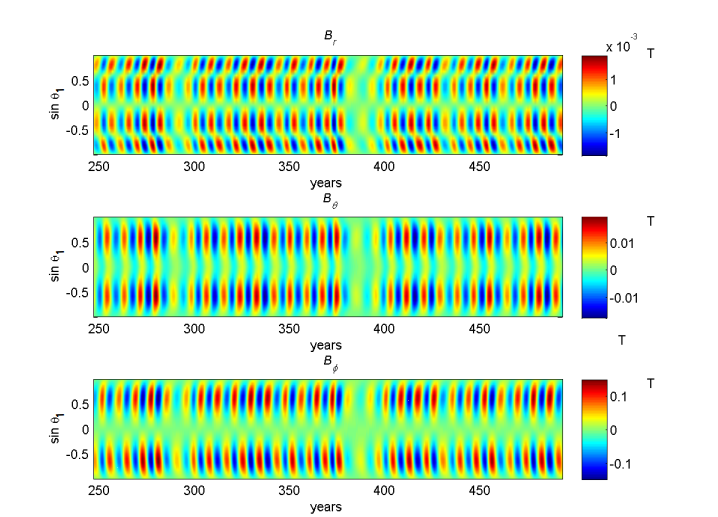}\hfill
    \caption{Magnetic field components of $\varepsilon$ Eridani as a function of time and  latitude at the surface just above the tachocline r=0.8$R_*$. The top and middle panels show the poloidal components  and  the bottom panel presents the toroidal component.}
  \label{fig:btime2}
\end{figure}

In Fig. \ref{fig:btime} and Fig. \ref{fig:btime2} we plot the components of the magnetic field derived from the model described.  The 
top and middle panels show the poloidal components  as function of latitude and time just below the surface, and just above the tachocline, respectively. The bottom panel presents the toroidal component as function of latitude and time at $r=0.98R_*$ and at $r=0.8R_*$. 

The large-scale magnetic field presents a topology similar to the solar one. In particular, the amplitude of $\langle B_\phi\rangle$ just above the tachocline is about 10 $\mu$T, and 0.1 T near the surface. Also, near the tachocline the toroidal field shows a shift of its maximum magnitude from middle latitudes towards 
the poles as the cycle advances, and the radial field also shows a poleward migration at high latitudes (Fig. \ref{fig:btime2}). Additionally, there is a 
phase lag between the maxima (and equivalently of the minima) of  $B_r$ and $B_\phi$ leading
to a negative correlation between them, consistent with a solar-type dynamo (see \cite{Charbonneau2010}).

From visual inspection, we see the polarity reversals in the toroidal components of the magnetic field every 4 years, approximately, and also that the amplitude of each component is modulated with a decadal periodic signal. This decadal variation is more evident in the radial component in the upper panel in Fig. \ref{fig:btime}, where we see consecutive maxima at 262 and 271.5 years or at 437 and 446.5 years. Furthermore, a time interval of very low activity, similar to grand minima, are present in all the magnetic field components between 286 and 300 years and between 383 and 400 years.

To quantify this variation, we search for a magnetic proxy derived from the simulated magnetic fields. To do so, we followed the well-known long-term analysis performed from solar observations (see \cite{Hathaway15} and references therein). For instance, several researchers consider the entire solar magnetic field on the solar disk to be a proxy for the generation of solar irradiance (\cite{Ball12}) or as the main proxy defining the whole picture of the solar activity as observed both
in time and in latitudes (\cite{Zharkova12}).
In fact, the solar activity cycle is reflected in both the poloidal and toroidal magnetic field components (\cite{Choudhuri2007,Charbonneau2010,Shepherd14}). In this sense, we define an activity index  $I_B$ in terms of the projection along the line of sight of the three components of the  magnetic field near the surface (Fig. \ref{fig:btime}), integrated over the stellar surface  (taking also into account the projection along the line of sight of the surface element):
\begin{equation}
I_B =\int_{-\pi }^{\pi }d\phi \int_{0}^{\pi}d\theta\,\,  F\left(x\right)\left\vert B_n\right\vert \sin \theta,
\label{eq.IB}
\end{equation}
\\
where $B_n=B_r (sin\theta cos\phi sin\beta+cos\theta_1 sin\beta)+ B_\theta (cos\theta sin\phi sin\beta -sin\theta cos\beta)+B_\phi (-sin\phi sin\beta)$; $x=\sin \beta \sin \theta \cos \phi + \cos \beta \cos \theta$, with $\beta$ the angle between the $z$ axis and the direction of observation and $\theta$ is the polar angle. Following \cite{Cr6} we take for $\varepsilon$ Eridani, $\beta\sim$ 30$^\textrm{o}$.
The function $F$, given by
\begin{equation}
F(x)=\Bigg\{ \begin{array}{lc c}
 x &\textrm{if\-} &x>0\\
  0 &\textrm{if\-}& x<0
\end{array}
\label{eq.f}
\end{equation}
takes account of the projection along the line of sight of the area element.

We remark that $I_B$  is not a unique activity indicator to perform this analysis. In particular, the Stokes $V$ profile is proportional to  this index and it can be closely related to the magnetic features observed in the stellar surface.

\begin{figure}
\includegraphics[width=\textwidth]{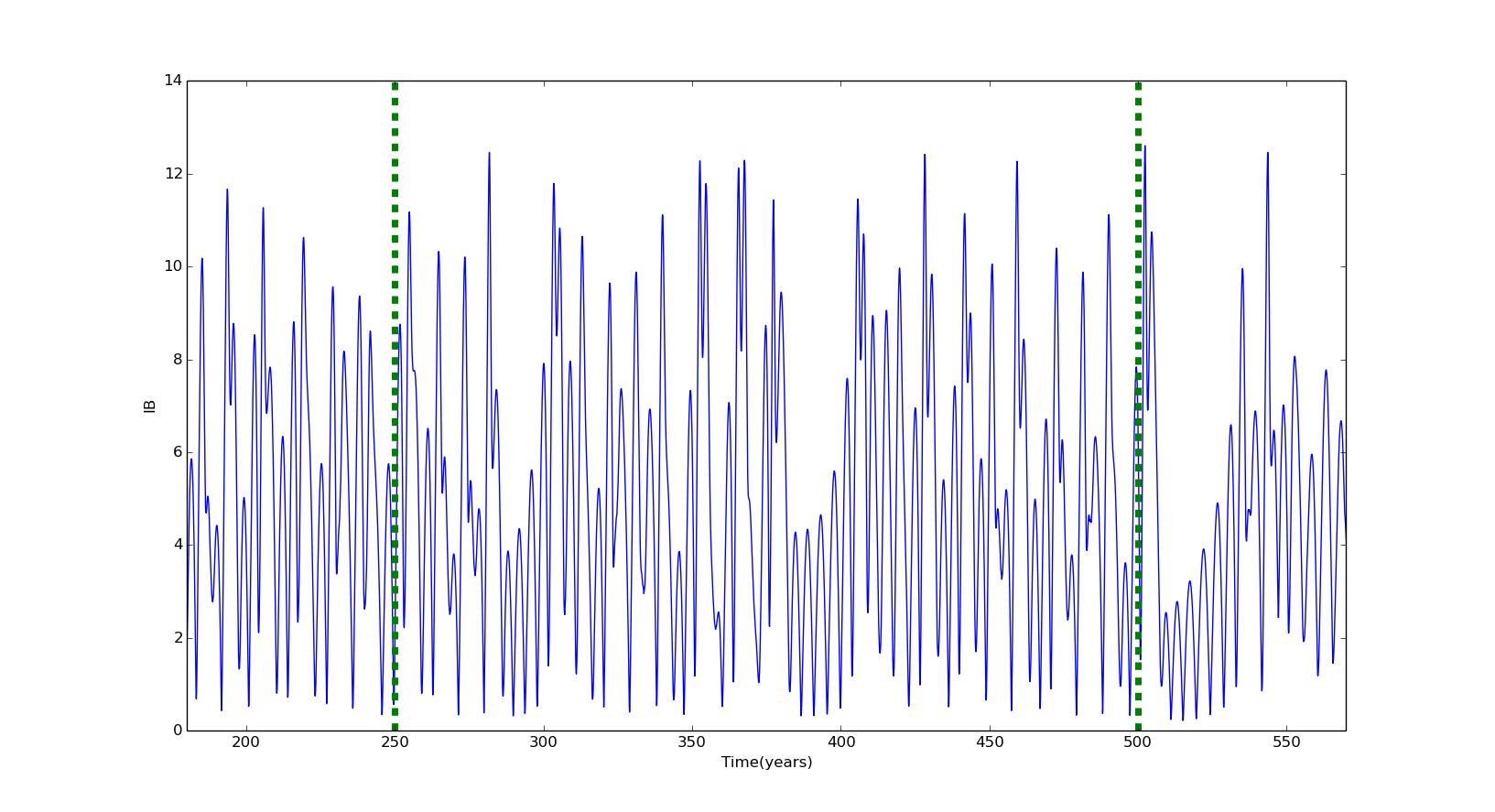}
\caption{$I_B$ index derived from the the magnetic field  projected on the line of sight (see Eq.\ref{eq.IB}) near the surface simulated by  the dynamo model for $\varepsilon$ Eridani. Dashed vertical lines indicate the interval time covered in  Fig. \ref{fig:btime} and Fig. \ref{fig:btime2} and Fig. \ref{fig:alfa}. }\label{fig.IBepseri}
\end{figure}

In Fig. \ref{fig.IBepseri} we plot the resulting time series of the $I_B$ index. We observe a short-term activity cycle modulated by a longer one and a broad minimum of $\sim$25 year length at $t=400$ yr, while the short activity cycle is still present.  Similarly, a decay of activity of only 13-year length is also observed between 286 and 299 yr.
Since this is a 2D model, we do not observe signs of rotational modulation in the time series. Nevertheless in real observations of activity proxies, short-term variations due to stellar rotation and instrumental noise are also present in the series.

To properly quantify the variations of the index, we analyzed the full $I_B$ series with the GLS periodogram (\cite{Zechmeister09}).

\begin{figure}
  \centering
    \includegraphics[width=\textwidth]{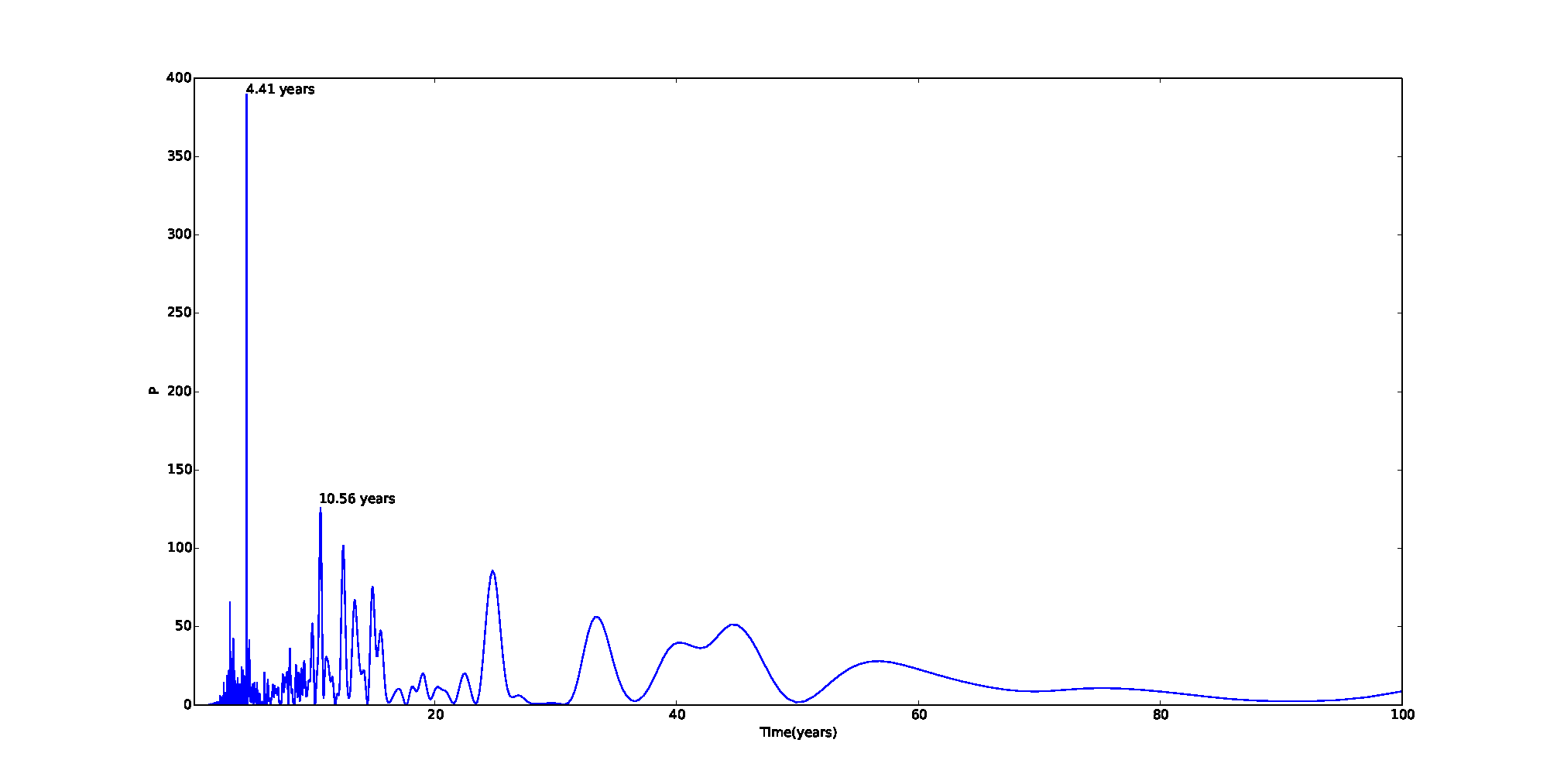}
  \caption{GLS periodogram obtained from the time-series plotted in Fig. \ref{fig.IBepseri}}
  \label{fig.periodogram}
\end{figure}

In Fig. \ref{fig.periodogram} we detected the most significant peak at 4.41 years and another significant peak at 10.56 years. All these periodic patterns are also observed in the $\alpha$-effect. As mentioned above the $\alpha$-effect is proportional to the cylindrical radial component of the vorticity of the flow. The contribution  $\alpha^{(0)}$ comes from the differential rotation, while the induced meridional flow contributes the part $\alpha^{(1)}$ (see Eq.\ref{eq.alf0} and Eq. \ref{eq.alf1}).
 
\begin{figure}
\centering
\includegraphics[width=1.1\textwidth]{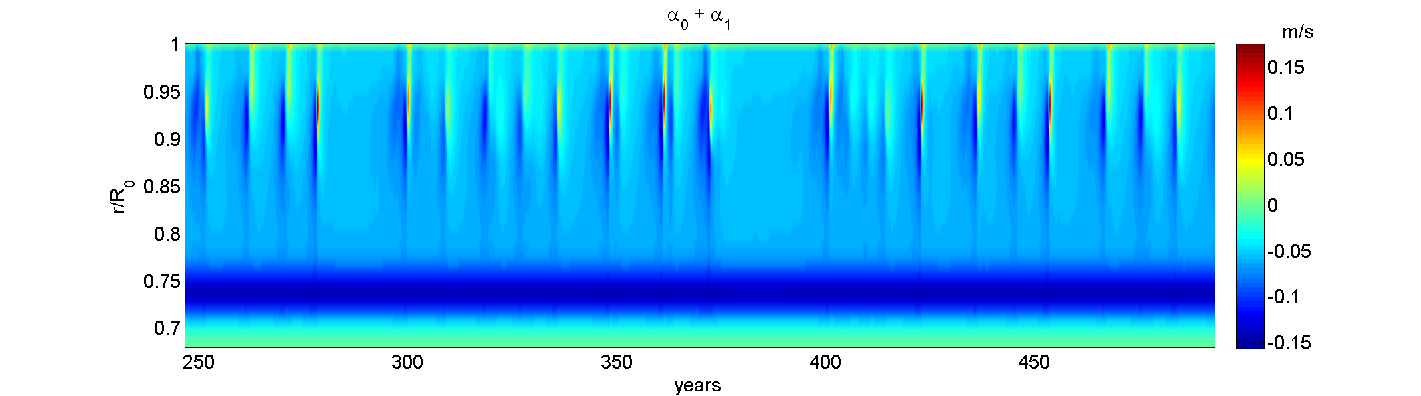}
  \caption{$\alpha^{(0)}$+$\alpha^{(1)}$ at latitude 30$^\textrm{o}$ vs. time}
  \label{fig:alfa}
\end{figure}

 In Fig. \ref{fig:alfa} we plot both components ($\alpha^{(0)}+\alpha^{(1)}$ ) of the $\alpha$ coefficient  in the stellar interior as function of time at latitude 30$^\textrm{o}$ where the first spots usually emerge during the solar cycle. The coefficient  $\alpha^{(0)}$ is constant in time (see Eq. \ref{eq.alf0}) and located mainly about the tachocline, while the $\alpha^{(1)}$-coefficient varies in time (see Eq. \ref{eq.alf1}) and has its higher magnitude near the surface. We remark that for $\varepsilon$ Eridani both coefficients are of the same order of magnitude. A decadal cyclic pattern, similar to the longer activity cycle,  is observed in the intensity of the  $\alpha$ coefficient, most evident in the stellar surface. Furthermore, in Fig. \ref{fig:alfa} this pattern is interrupted  in the intervals between 281 and 298 years and between 374 and 400 years, this long decay of the $\alpha$ coefficient seems to be correlated with the broad minimum also observed in the $I_B$-series at 286-299 years the intervals  and 375-400 years respectively .

\begin{figure}
  \centering
    \includegraphics[width=0.45\textwidth]{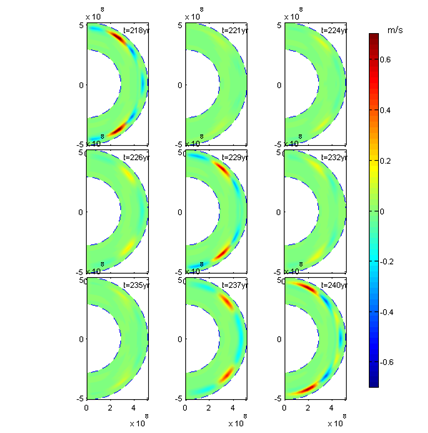}\hfill 
    \includegraphics[width=0.455\textwidth]{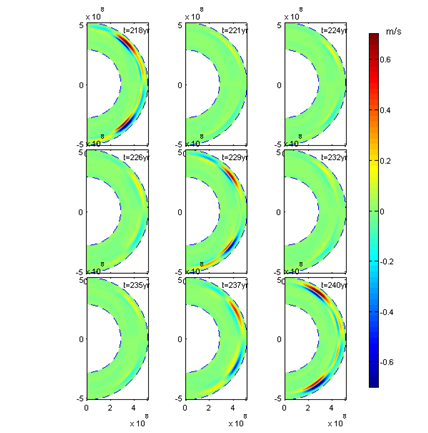}
  \caption{ Radial  (left) and Poloidal (right) component of the meridional velocity field  vs. time. The nine sequential boxes cover 22-year span with a 2.5-yr sampling. The dashed blue lines indicate the boundary radii of the model ($R_{ext}=R_\star$ and $R_{int}=0.55R_\star$)}\label{fig.urtit}
\end{figure}

In Fig. \ref{fig.urtit}  we plot the radial and  poloidal (i.e. $\theta$) components of the induced meridional velocity  during 22 years, with a a 2.5-yr sampling. We selected an arbitrary time, t=218 yr, after a few hundred years of simulated time. Both components of the meridional flow are not null near the surface and each pattern presents a decadal cyclical variation . 

In particular, near the maxima ($t=218, 229, 240$), the radial component indicates downstream flows near the equator and at low latitudes, and upstream flows at middle latitudes in both hemispheres, a behaviour opposite to the solar case. 

Given both the radial and poloidal components, we reconstruct a unique cell of  clockwise (counterclockwise) circulation in the northern (southern) hemisphere,  contrary to the solar case (e.g. \cite{Gonzalez06,Hathaway2010,Sraibman19}). Furthermore, during maxima, a second cell is also present near the poles with counterclockwise and clockwise circulation at northern and southern hemispheres, respectively. 

\section{Discussion}\label{sec.discuss}

During the last decade, 
high-cadence observations, which allow the detection of short activity cycles (e.g. \cite{Metcalfe10,Mittag19}), the assembly of different stellar activity measurements, which allows to build long  reliable activity series (e.g. \cite{Metcalfe13,Egeland17,Soon19}) and advances in time-series techniques (e. g. \cite{Zechmeister09,Mortier15,Olspert18}) have led to a revision  of the magnetic activity in solar-type stars. 

Many of these works have paid special attention to the relation  between stellar activity cycles and the stellar rotation rate (e.g. \cite{BohmVitense07,Olah09, Barnes10,Metcalfe16}). One of the most attractive points is that, given a stellar rotation period, the activity cycle period could be distributed in two distinct branches, as first invoked in \cite{Brandenburg98}.

The analysis and discussion on this bimodal distribution  have been addressed either from observations, dynamo models or statistical analysis (e.g. \cite{Metcalfe16,Metcalfe17, See16, Brandenburg17,Strugarek17,Strugarek18, Boro18}).

In this work, we take a new approach to this discussion.
We built a non-linear axisymmetric dynamo model of a single K2V star $\varepsilon$ Eridani. 
The fact that $\varepsilon$ Eridani has been widely observed by different programs allows \cite{Metcalfe13} to build one  of the most reliable and extended registry of stellar activity.  The analysis of this series reveals two simultaneous activity cycles of $\sim$3- and $\sim$13-year period and a broad minimum of activity of 7 year-length which resembles  the solar Maunder Minimum. All these features make $\varepsilon$ Eridani a unique star to analyze both simultaneous activity cycles from the point of the dynamo theory. Furthermore,  several stellar parameters of $\varepsilon$ Eridani  are well-determined from accurate observations. The stellar surface differential rotation derived  in \cite{Cr6} and the internal structure obtained from seismic analysis in \cite{Gai08}, both from MOST photometry, allow us to obtain precise parameters for our dynamo model, based on the model developed in \cite{Sraibman19} for the Sun.

We derived the large-scale magnetic
field geometry of the young active K dwarf $\varepsilon$ Eridani over decades. To compare our simulations with  magnetic field observations, we should remark that the \textit{visible} magnetic field of our model is  composed by the simulated $B_r$ and $B_\theta$ components at the surface as the model does not include the emergence mechanism of the toroidal component to the stellar surface. In this sense both components projected along the line of sight and averaged over the stellar surface reaches maxima between 0.022 and 0.027 T and also exhibits a short cyclic variation, in agreement with \cite{Lehmann15}. Furthermore, the mean $\langle B_r\rangle\sim 6.8\times 10^{-3}$ T and $\langle B_\theta\rangle\sim 2.8\times 10^{-3}$ T which are of the same of order of the magnetic field derived  by \cite{Jeffers14} from ZDI observations.

Nevertheless,  to properly  compare the evolution of the simulated magnetic fields with the Mount Wilson series built in \cite{Metcalfe13}, we defined an index of activity related to the line-of-sight component of the total magnetic field near the surface, integrated in the full disc. The resulting time-series, analyzed with a periodogram technique, reveals a short activity cycle $\sim$4-year period coexisting with  a longer one of  $\sim$11 years, in agreement with \cite{Metcalfe13}. Furthermore, the model also reproduces low stellar activity phases, similar to the one registered between 1985  and 1992 by the authors. Nevertheless, the simulated short activity cycle is still present during the activity minima, but it seems to be absent in the observations.

 
Our model reveals that the short activity cycle could be reproduced with an axisymmetric dynamo model without feedback, outlined in \cite{Sraibman17} and in \cite{Sraibman16} for the solar case. This short cycle associated to the toroidal  magnetic fields reversals is driven by the differential rotation. In the present work we found that the long activity cycle and the activity minima are related to the meridional mass flows induced by the Lorentz force. This last phenomenon is evident near the stellar surface in the simulated meridional velocity component and the $\alpha$-effect. 

While \cite{BohmVitense07}  assigns the long cycle to a dynamo operating near the surface and the short cycle to a second co-existing dynamo near the tachocline and \cite{Brandenburg17} concluded that all stars younger than 2.3 Gyr sustain both dynamos, we found that a single non-linear axisymmetric dynamo  could reproduce both long and short activity cycles. 
In agreement with the former works our  model reveals that the long activity cycle of the young star $\varepsilon$ Eridani (0.8 Gyr) is  evident  in the  stellar surface (See Fig. \ref{fig:alfa} and Fig. \ref{fig.urtit}) and the short activity cycle  in the bottom of the convection zone.

Furthermore, our simulations on this particular star reproduce the field magnitudes obtained from a global MHD model developed in \cite{Strugarek18} for a grid of stellar parameters. Also the same study suggests that stars with Rossby number $R_o$ between 0.25 and 1 could present a short and long coexisting activity cycles, and $\varepsilon$ Eridani is in this range with a $R_o=0.48$ (\cite{See16}).

Therefore we conclude that a complete non-linear axisymmetric dynamo model, as the one  presented in this work  could be a reasonable approximation of a realistic dynamics of the stellar magnetic field in solar-type stars. Given accurate stellar parameters, this model is a useful tool to analyze their long-term activity. In particular, it could reproduce both the long and short activity cycles in $\varepsilon$ Eridani with results similar to those of previous works.


\section*{Acknowledgements}
We acknowledge the Consejo Nacional de Investigaciones Cient\'\i ficas y T\'ecnicas
(CONICET) and the University of Buenos Aires for institutional support.



\section*{Data Availability}
The data underlying this article will be shared on reasonable request to the corresponding author.
\bibliographystyle{mnras}
\bibliography{biblio}
\end{document}